\documentclass[reqno,12pt]{article}
\usepackage{amsmath,amsfonts,amssymb,amsthm,amstext,amscd,eucal,xcolor}
\usepackage{color}
\usepackage[all]{xy}
\usepackage{cite,hyperref}
\usepackage{graphicx}
\usepackage{epsfig,epstopdf}
\usepackage[active]{srcltx}
\makeatletter \@addtoreset{equation}{section}

\makeatletter\renewcommand\section{\@startsection {section}{1}{\z@}%
                                   {-3.5ex \@plus -1ex \@minus -.2ex}
                                   {2.3ex \@plus.2ex}%
                                   {\normalfont\large\bfseries}}
\renewcommand\subsection{\@startsection{subsection}{2}{\z@}%
                                     {-3.25ex\@plus -1ex \@minus -.2ex}%
                                     {1.5ex \@plus .2ex}%
                                     {\normalfont\bfseries}}

\newcommand{\be}{\begin{equation}}
\newcommand{\ee}{\end{equation}}
\newcommand{\bea}{\begin{eqnarray}}
\newcommand{\eea}{\end{eqnarray}}
\newcommand{\bse}{\begin{subequations}}
\newcommand{\ese}{\end{subequations}}
\newcommand{\beqa}{\begin{eqnarray}}
\newcommand{\eeqa}{\end{eqnarray}}
\newcommand{\beqar}{\begin{eqnarray*}}
\newcommand{\eeqar}{\end{eqnarray*}}
\newcommand{\bi}{\begin{itemize}}
\newcommand{\ei}{\end{itemize}}
\newcommand{\bn}{\begin{enumerate}}
\newcommand{\en}{\end{enumerate}}

\newcommand{\ba}{\begin{array}}
\newcommand{\ea}{\end{array}}
\newcommand{\bc}{\begin{center}}
\newcommand{\ec}{\end{center}}

\definecolor{darkgreen}{rgb}{0,0.3,0}
\definecolor{darkblue}{rgb}{0,0,0.3}
\definecolor{darkred}{rgb}{0.7,0,0}

\newcommand{\NC}{noncommutative }
\newcommand{\email}[1]{\footnote{E-mail: \href{mailto:#1}{#1}}}

\makeatother

\begin{document}

\title{\textbf{\Large
{One-loop $\mathbf{\beta}$ function of noncommutative scalar $QED_{4}$}}}

\author{\textbf{M.~Ghasemkhani}$^{1}$\email{m$_{_{-}}$ghasemkhani@sbu.ac.ir} \textbf{,
R.~Bufalo}$^{2}$\email{rodrigo.bufalo@dfi.ufla.br}\textbf{, V. Rahmanpour}$^{1}$\email{v.rahmanpour@mail.sbu.ac.ir
}
\textbf{and E. Nouri}$^{1}$\email{el.nouri@mail.sbu.ac.ir}\\\\
 \textit{$^{1}$} \textit{\small Department of Physics, Shahid
Beheshti University,} \\
 \textit{\small G.C., Evin, Tehran 19839, Iran} \\
\textit{$^{2}$ \small Departamento de F\'{i}sica, Universidade Federal de Lavras,}\\
\textit{\small Caixa Postal 3037, 37200-000 Lavras, MG, Brazil} \\
}
\maketitle
\begin{abstract}
In this paper we consider the $\beta$ function at one-loop approximation for noncommutative scalar QED.
The renormalization of the full theory, including the basic vertices, and the renormalization group equation are fully established.
Next, the complete set of the one-loop diagrams corresponding to the first-order radiative corrections to the basic functions is considered: gauge, charged scalar and ghost fields self-energies, and three- and four-point vertex functions $\left<\phi^\dag \phi A\right>$, $\left<\phi^\dag \phi A A\right>$, and $\left<\phi^\dag \phi\phi^\dag \phi\right>$, respectively. We pay special attention to the noncommutative contributions to the renormalization constants.
To conclude, the one-loop $\beta$ function of noncommutative scalar QED is then computed and comparison to known results is presented.

\end{abstract}
\begin{flushleft}
\par\end{flushleft}


\newpage
\tableofcontents


\section{Introduction}
\label{sec:1}
Perturbative gauge theories are the most successful models for the description of the fundamental interactions in nature, where the coupling constant is small. One of the most successful of these gauge theories  is QED, a theory that describes the interaction of fermionic charged particles with electromagnetic field. In the QED framework, the one-loop analysis for the physical quantities, such as the electron anomalous magnetic moment and the Lamb shift effect in the energy levels of a hydrogen atom, gives us theoretical predictions which are in excellent agreement with experimental data \cite{feynman}.
In fact, the most accurate low-energy measurement of the electromagnetic fine structure constant (the strength of the electromagnetic interaction) originates from the electron anomalous magnetic moment, which was measured precisely using a single electron caught in a Penning trap \cite{odom}.

As we know, some physical quantities in a field theory do vary with the related energy scale $\mu$ in which it is considered; for example, the value of the fine structure constant runs with growing energy scale $\mu$. To account for such behavior of all physical quantities, the renormalization group program can suitably be used so that we can define consistently the physical outcomes for a given field theory \cite{weinberg}.
In particular, in this context, two functions play a major role: they are the beta and gamma functions that  measure the running of a coupling constant on the energy scale $\mu$ and the anomalous dimension of correlators, respectively.
For instance, perturbative analysis for QED shows that the one-loop $\beta$ function is found as
$\beta(e)_{_{\tiny{\rm{QED}}}}=\frac{e^{3}}{12\pi^{2}}$ (with $N_{F}=1$) \cite{weinberg}. This physically means that the coupling increases with an increasing energy scale so that QED becomes strongly coupled at high energies.

Aside from this, to describe the dynamics of spinless charged fields interacting with photons, a suitable framework is the scalar quantum electrodynamics (SQED), and it is known that its one-loop beta function is expressed by $\beta(e)_{_{\tiny{\rm{SQED}}}}=\frac{e^{3}}{48\pi^{2}}$ (with $N_{B}=1$) \cite{crednicki}.
By a simple comparison of this result to that one from QED, it is easily realized that the signs of the both one-loop $\beta$ functions are positive, showing thus that QED and SQED are infrared free at one-loop order and moreover that the identity in Eq.~\eqref{eq:0.001} holds,
 \begin{equation}
\beta_{_{\tiny{\rm{QED}}}}=4\beta_{_{\tiny{\rm{SQED}}}},\label{eq:0.001}
\end{equation}
 in which we understand that the coefficient 4 comes from the trace over the gamma matrices due to fermionic loops. However, it is important to emphasize that the relation \eqref{eq:0.001} is valid at one-loop order only and in general it is not satisfied by higher-loop perturbative order. More details on this issue can be found in Refs.~\cite{Chetyrkin-1980,Chetyrkin-1981}.
 We can then observe that the \emph{spin}, as an additional degree of freedom for interacting charged fields in QED, only changes the \emph{intensity} of the one-loop $\beta$-function and not its \emph{sign}.
 Besides, if we consider additional degrees of freedom for spinning charged fields such as color, e.g., in quark matter, then once again it will be observed that the sign of the one-loop $\beta$ function for the model describing the interaction of quarks with photons, does not change.
On the other hand,  the structure of the $\beta$ function for the interaction of quarks with gluons in QCD does basically change in comparison to the latter case, since the internal gauge symmetry now is completely changed \cite{weinberg}.

 As is known the study of noncommutative gauge theories throughout the years has uncovered several interesting physical properties \cite{ref16, ref23, Amelino-Camelia:2013fxa}, so it would be valuable to investigate whether the relation \eqref{eq:0.001} is also satisfied for the noncommutative setup. In other words, we wish to examine whether or not adding a new degree of freedom to charged fields in a noncommutative spacetime only changes the intensity of the $\beta$ function.
 Furthermore, it is interesting to understand how the noncommutativity affects the structure of the $\beta$ functions for these QED and SQED theories. The answer to the second question in the case of noncommutative QED at one-loop order has already been studied in Refs. \cite{Martin-1999,hayakawa-1} and is given by
\begin{align}
\beta(e)_{\rm{NC-QED}} = -\frac{e^{3}}{16\pi^{2}}\left(\frac{22}{3}-\frac{4N_{F}}{3}\right).
\label{ncqed-beta}
\end{align}
The obtained result is independent of the noncommutativity parameter $\theta$, but its structure is completely different from its commutative counterpart. Indeed, the one-loop contributions of the relevant graphs to this $\beta$ function arise only from the planar parts, while the nonplanar parts are all finite. Although the nonplanar part of the respective diagrams does not contribute to the $\beta$ function at one-loop order, the noncommutative effects are actually presented by means of the new couplings engendered by the noncommutativity of spacetime coordinates.
The structure of Eq.~\eqref{ncqed-beta} is similar to those from non-Abelian gauge theories, in particular the $SU(2)$ gauge theory.
Besides, we see that for $N_{F}=0$ the theory reduces to a pure gauge part with a negative $\beta$ function which is asymptotically free.

However, the one-loop $\beta$ function of the noncommutative SQED has not yet been computed, and the exact form of the one-loop $\beta$ function for NC-SQED could be naively obtained through the relation \eqref{ncqed-beta} and the aforementioned considerations in Eq.~\eqref{eq:0.001}. To this end, we notice that the contribution $\frac{22}{3}$ originates just from the pure gauge sector; hence, this part should also be present in NC-SQED, while the second term, including the matter sector, similar to the commutative case would have a different coefficient (due to its spinless structure). Consequently, we find
\begin{align}
\beta(e)_{\rm{NC-SQED}} = -\frac{e^{3}}{16\pi^{2}}\left(\frac{22}{3}-\frac{N_{B}}{3}\right).
\label{ncsqed-beta}
\end{align}
One of the main goals in this paper is to correctly establish the above result in a more detailed analysis, by considering the interaction among the spinless matter and gauge fields in the noncommutative SQED.
Some general discussions on the renormalization of SQED have been considered before \cite{Blaschke:2009rb,Blaschke:2005dv,Martin:2006gw}, but with a different scope than ours.
More importantly, renormalization of noncommutative gauge theories must be treated carefully and is still the subject of analysis \cite{grosse-1,grosse-2,D'Ascanio:2016asa}.
Moreover, it is worth noticing the similarity between the $\beta$ functions of the noncommutative Abelian gauge theory and commutative non-Abelian gauge theory presented above. Hence, we will discuss throughout the paper to what extent this similarity holds and also the role played by the spin of the matter field in the $\beta$ function of these theories.

Therefore, the organization of this paper is as follows. In Sec.~\ref{sec:2}, after introducing the Lagrangian of the NC-SQED model and its invariance under the Becchi-Rouet-Stora-Tyutin (BRST) Slavnov transformations, we present the relevant Feynman rules used in our one-loop calculations. Then in Sec.~\ref{sec:3}, we discuss the renormalization procedure for the full Lagrangian density and in particular specify the related renormalization constants for the matter, gauge, and interaction terms. In addition, we show that these renormalization constants satisfy in the Slavnov-Taylor identities. Next, in Sec.~\ref{sec:4}, the one-loop analysis for the self-energy of the gauge, scalar, and ghost field is performed, which yields us the one-loop renormalization constants for the related fields. Also, the radiative correction to three- and four-point vertices, including two scalars and one photon, two scalars and two photons  and four scalars, respectively, is carried out in detail, which gives us four renormalization constants corresponding to those three vertices. Based on the obtained results for the renormalization constants, the one-loop $\beta$ function and also the anomalous dimensions of the scalar, gauge, and ghosts fields are presented and discussed in Sec.~\ref{sec:5}. Finally, Sec. 6 is dedicated to the concluding remarks.

\section{Model}

\label{sec:2}

The gauge-fixed dynamics of noncommutative scalar QED is described by the Lagrangian
density
\begin{align}
{\cal {L}}=({\cal{D}}^{\mu}\phi)^{\dag}\star ({\cal{D}}_{\mu}\phi) -m^{2}\phi^{\dag}\star\phi-\frac{1}{4}F_{\mu\nu}\star F^{\mu\nu}+{\cal{L}}_{\phi^4}+{\cal {L}}_{g.f}+{\cal {L}}_{gh}, \label{eq:0.1}
\end{align}
where the covariant derivative is defined as ${\cal{D}}_{\mu}\phi=\partial_{\mu}\phi+ieA_{\mu}\star\phi$ and the field strength tensor $F_{\mu\nu }=\partial_\mu A_\nu -\partial_\nu A_\mu +ie \left[A_\mu,A_\nu\right]_\star$, such that $\left[~,~\right]_\star$ is the Moyal bracket.
Also, the Moyal star product between the functions $f(x)$ and $g(x)$ is described as
\begin{equation}
f\left(x\right)\star g\left(x\right)=f\left(x\right)\exp\bigg(\frac{i}{2}
\theta ^{\mu\nu} \overleftarrow{\partial_{\mu}}~\overrightarrow{\partial_{\nu}}\bigg)g\left(x\right).
\end{equation}
Additionally, a noncommutative counterpart of the quartic interaction among the charged scalar fields is necessary in order to assure the theory's full renormalizability \cite{Arefeva:2000uu},
\begin{equation}
\mathcal{L}_{\phi^4}=-\frac{1}{4}\bigg[\lambda_1(\phi^{\dagger}*\phi)^2+\lambda_2(\phi^{\dagger}*\phi^{\dagger}*\phi*\phi)\bigg]
\end{equation}
where $\lambda_1$ and $\lambda_2$ are coupling constants and play an important part regarding the renormalizability and infrared behavior (UV/IR mixing).
The gauge-fixing term is chosen on the Lorenz condition
\begin{equation}
{\cal {L}}_{g.f} =\frac{\xi}{2}B\star B+B\star \partial_\mu A^\mu,
\end{equation}
and $B$ is the Nakanishi-Lautrup auxiliary field, while the ghost contribution reads
\begin{equation}
{\cal {L}}_{gh} =\partial_\mu \bar{c} \star D^\mu c,
\end{equation}
where we have defined $D_{\mu} \bullet=\partial_{\mu} \bullet+ie \left[A_{\mu},\bullet\right]_\star$.
The above Lagrangian is invariant under BRST Slavnov transformations   \cite{Martin:2000dq}
\begin{align}
sA_\mu = D_\mu c, \quad s \phi =ie c\star \phi , \quad sc = ie c\star c , \quad s\bar{c} = -B, \quad sB=0.
\end{align}
We will show next that under discrete symmetries SQED in the noncommutative setup has a similar behavior as in the NC-QED \cite{SheikhJabbari:2000vi}; this can be seen from the Lagrangian density \eqref{eq:0.1} as follows
\begin{itemize}
  \item [(\emph{i})]~\emph{Parity} \newline
  \newline
  The transformation of the pure gauge sector under parity has been carried out in Ref. \cite{SheikhJabbari:2000vi}, where it was shown that the respective pure gauge terms in the NC-QED model are invariant under parity without any change in the sign of $\theta$. To study the behavior of the matter part under parity, we consider the relevant term given by
  \begin{align}
{\cal {L}}_{matter}=ie\left( \partial^{\mu}\phi^{\dag} \star A_{\mu} \star \phi- \phi^{\dag}\star A_{\mu} \star \partial^{\mu}\phi \right)
+e^2 \phi^{\dag}\star A^{\mu}\star A_{\mu}\star\phi+{\cal{L}}_{\phi^4}.\label{matter}
  \end{align}
  Under parity, the complex scalar field transforms as $\phi_{_{P}}(x')=e^{i\alpha}\phi(x)$, in which the phase $\alpha$ is arbitrary for a free field \cite{pokorski}. Besides, the gauge field components change as $A^{0}_{_{P}}(x')=A^{0}(x)$ and $\textbf{A}_{_{P}}(x')=-\textbf{A}(x)$ so that we easily conclude ${\cal {L}}_{matter}$ does not change under the parity and thus NC-SQED is parity invariant.
   \item [(\emph{ii})]~\emph{Charge conjugation}\newline
  \newline
   Under charge conjugation, we have $A_{\mu}^{c}(x)=-A_{\mu}(x)$ and $\phi^{c}(x)=e^{i\eta}\phi^{\dagger}(x)$, where $\eta$ is an arbitrary phase parameter \cite{pokorski}. Taking into account these changes as well as $\theta\rightarrow -\theta$, we observe that the first term of ${\cal {L}}_{matter}$ goes to the second term and vice versa. Furthermore, it is easily seen that the third and fourth terms in \eqref{matter} do not change, and then by considering the invariance of the gauge part as in Ref. \cite{SheikhJabbari:2000vi}, we consequently deduce that NC-SQED is invariant under charge conjugation transformation (\emph{with} $\theta\rightarrow -\theta$).
   \item [(\emph{iii})]~\emph{Time reversal}\newline
  \newline
     The time reversal operator acts on the gauge and the scalar fields as $A^{0}_{_{T}}(x')=A^{0}(x)$, $\textbf{A}_{_{T}}(x')=-\textbf{A}(x)$, and $\phi_{_{T}}(x')=e^{i\zeta}\phi^{\dagger}(x)$, respectively. Similar to the aforementioned discussion on charge conjugation, once again, we see that the first and the second terms of Eq.~\eqref{matter} transform to each other and the third and fourth terms remain unchanged, if we assume that $\theta\rightarrow -\theta$. Together with the invariance of the gauge part, we realize that NC-SQED is time reversal invariant (\emph{with} $\theta\rightarrow -\theta$).
\end{itemize}
Hence, from the above discussion on the discrete symmetries of NC-SQED, we reach a result which holds also for NC-QED and was already found in Ref.~\cite{SheikhJabbari:2000vi}. Actually, it is understood that NC-SQED is parity invariant without assumption $\theta\rightarrow -\theta$, and hence this is the same as the commutative setup. However, in order to have charge conjugation and time reversal invariance, we should take into account \emph{additional}
transformation for the noncommtativity parameter as $\theta\rightarrow -\theta$. Hence, NC-SQED, similar to NC-QED, is a \emph{CP}-violating model, but is \emph{CPT} invariant.

Next, based on the gauge-fixed Lagrangian  \eqref{eq:0.1}, one can integrate over the auxiliary field $B$, so we can derive all the necessary Feynman rules for the propagators and vertex functions:
\begin{itemize}
\item Gauge field propagator (at Feynman gauge $\xi =1$):
\begin{equation}
D_{\mu\nu}(k)=\frac{-ig_{\mu\nu}}{k^{2}}.
\end{equation}
\item Scalar field propagator:
\begin{equation}
S(k)=\frac{i}{k^{2}-m^{2}}.
\end{equation}
\item Ghost field propagator:
\begin{equation}
D(k)=\frac{i}{k^{2}}.
\end{equation}
\item Three-point vertex $\left<\phi^\dag \phi A\right>$:
\begin{equation}
\Gamma^{\mu}(p,p')=-ie(p+p')^{\mu}e^{\frac{i}{2}p\wedge p'}.
\end{equation}
\item Four-point vertex  $\left<\phi^\dag \phi A A\right>$:
\begin{equation}
\Upsilon^{\mu\nu}(p,p',k,k')=2ie^{2}g^{\mu\nu}e^{\frac{i}{2}p\wedge p'}\cos(\frac{k\wedge k'}{2}).
\end{equation}
\item Three-point vertex $\left<\bar{c}c A\right>$:
\begin{equation}
\Psi^{\mu}(p,p')=2ep^{\mu}\sin(\frac{p\wedge p'}{2}).
\end{equation}
\item Cubic gauge vertex $\left<AA A\right>$:
\begin{equation}
\Omega^{\mu\nu\rho}(p_{1},p_{2},p_{3})=2e\sin(\frac{p_{1}\wedge p_{2}}{2})\bigg[
g^{\mu\nu}(p_{1}-p_{2})^{\rho}+g^{\nu\rho}(p_{2}-p_{3})^{\mu}+g^{\rho\mu}(p_{3}-p_{1})^{\nu}
\bigg].
\end{equation}
\item Quartic gauge vertex $\left<AAA A\right>$:
\begin{align}
\Delta^{\mu\nu\rho\sigma}(p_{1},p_{2},p_{3},p_{4})&=4ie^{2}\bigg[
(g^{\mu\rho}g^{\nu\sigma}-g^{\mu\sigma}g^{\nu\rho})
\sin(\frac{p_{1}\wedge p_{2}}{2})\sin(\frac{p_{3}\wedge p_{4}}{2})\nonumber\\
&+(g^{\mu\sigma}g^{\nu\rho}-g^{\mu\nu}g^{\rho\sigma})\sin(\frac{p_{3}\wedge p_{1}}{2})\sin(\frac{p_{2}\wedge p_{4}}{2})\nonumber\\
&+(g^{\mu\nu}g^{\rho\sigma}-g^{\mu\rho}g^{\nu\sigma})\sin(\frac{p_{1}\wedge p_{4}}{2})\sin(\frac{p_{2}\wedge p_{3}}{2})
\bigg].
\end{align}
\item Four-scalar vertex $\left<\phi^\dag \phi\phi^\dag \phi\right>$:
\begin{align}
i \Gamma(p_{1},p_{2},p_{3},p_{4}) =  \lambda_1\cos{\big(\frac{p_1\wedge p_2+p_3\wedge
p_4}{2}\big)}+\lambda_2 \cos{\big(\frac{p_1 \wedge
p_3}{2}\big)}\cos{\big(\frac{p_2 \wedge p_4}{2}\big)}.
\end{align}
\end{itemize}
Here, $p\wedge k=\theta^{\mu\nu}p_{\mu}k_{\nu}$, and the symbol $\left<\cdots\right>$ indicates the vacuum expectation value of the time-ordered operators, corresponding to the $n$-point function of the fields.
Before proceeding with computing the one-loop corrections to the basic functions, we shall now establish the renormalization of the given theory.
\section{Renormalization}
\label{sec:3}
We present the multiplicative renormalization of the NC-SQED. The bare and renormalized fields are related by means of
\begin{equation}
A^{(0)}_\mu= \sqrt{Z_{_{3}}}~A_\mu, \quad  \phi^{(0)}= \sqrt{Z_{_{2}}}~\phi, \quad c^{(0)}= \sqrt{\widetilde{Z}_{_{3}}}~ c.
\end{equation}
Now, we rewrite the Lagrangian \eqref{eq:0.1} in terms of the renormalized fields so that we find explicitly for the interaction part
\begin{align}
{\cal {L}}_{\rm{int}}^{(r)}&	=ie Z_{_{1}} \left( \partial^{\mu}\phi^{\dag} \star A_{\mu} \star \phi- \phi^{\dag}\star A_{\mu} \star \partial^{\mu}\phi \right)  +e^2 Z_{_{4}} \phi^{\dag}\star A^{\mu}\star A_{\mu}\star\phi  \nonumber \\
& -ie Z_{_{3A}} \partial_\mu A_\nu   \star   \left[A^\mu,A^\nu\right]_\star
+\frac{e^2}{4} Z_{_{4A}} \left[A_\mu,A_\nu\right]_\star \star   \left[A^\mu,A^\nu\right]_\star \nonumber \\
&-\frac{\lambda_1}{4}Z_{\lambda_1}(\phi^{\dagger}*\phi)^2-\frac{\lambda_2}{4}Z_{\lambda_2}(\phi^{\dagger}*\phi^{\dagger}*\phi*\phi) +ie \widetilde{Z}_{_{1}} \partial^\mu \bar{c} \star \left[A_{\mu},c\right]_\star;
\end{align}
we can then introduce the counterterms as usual $Z_i = 1+\delta_i$.
Notice that the renormalization constant $Z_{_{1}}$ is related to the vertex $\left<\phi^\dag \phi A\right>$, $Z_{_{3A}}$ is related to the vertex $\left<AAA\right>$, $\widetilde{Z}_{_{1}}$ is related to the vertex $\left< \bar{c}cA\right>$, $Z_{_{4}}$ is related to the vertex $\left<\phi^\dag \phi A A\right>$, $Z_{_{4A}}$ is related to the vertex $\left<AA A A\right>$, and $Z_{\lambda_1}$ and $Z_{\lambda_2}$ are related to the two terms of the vertex $\left<\phi^\dag \phi\phi^\dag \phi\right>$.

Moreover, the gauge invariance, expressed in terms of the Slavnov-Taylor identities, assures the universality of the (gauge) coupling by renormalization, provided the following identities hold:
\begin{equation}
\frac{Z_{_{4A}}}{Z_{_{3A}}}= \frac{Z_{_{3A}}}{Z_{_{3}}} = \frac{\widetilde{Z}_{_{1}}}{ \widetilde{Z}_{_{3}}}
= \frac{Z_{_{1}}}{ Z_{_{2}}} = \frac{Z_{_{4}}}{Z_{_{1}}}. \label{STidentity}
\end{equation}

Since we are interested in computing the basic functions $\left<\phi^\dag \phi \right>$, $\left<A A\right>$, $\left<\bar{c} c\right>$, $\left<\phi^\dag \phi A\right>$, $\left<\phi^\dag \phi AA\right>$, and $\left<\phi^\dag \phi \phi^\dag \phi\right>$, we shall find by this analysis the respective renormalization constants $Z_{2}$, $Z_{3}$, $\widetilde{Z}_{3}$, $Z_{1}$, $Z_{\lambda_1}$, and $Z_{\lambda_2}$ so that in this case the renormalized coupling constants can be determined as $e^0= Z_e e$ and $\lambda^0_{1,2}= Z_2^{-2}Z_{\lambda_{1,2}} \lambda_{1,2} $, where we have defined conveniently $Z_e= Z_{_{1}} Z_{_{3}}^{-1/2} Z_{_{2}}^{-1} $ or $Z_e=  Z_{_{4}}^{1/2}Z_{_{3}}^{-1/2} Z_{_{2}}^{-1/2}$, where further relations for the renormalization constant $Z_e$ can be obtained by making use of the relations \eqref{STidentity}.
We note that, however, by determining these six renormalization constants, the remaining three are immediately determined by means of the gauge identities  \eqref{STidentity}. However, the one-loop analysis of the quantum corrections to the cubic and quartic gauge vertices, which yields $Z_{_{3A}}$ and $Z_{_{4A}}$, has already been presented in Ref. \cite{charneski}.

To conclude, one can work out the renormalization group equation for the renormalized Green function, establishing the invariance of the observables under changes of the renormalization scale $\mu$. In particular, let us consider the two-point function for the gauge field
\begin{equation}
\left[\mu \frac{\partial}{\partial \mu} +\beta (e)\frac{\partial}{\partial e}+\beta (\lambda)\frac{\partial}{\partial \lambda} - 2   \gamma_{_{A}} (e)\right] \Gamma^{\mu\nu}_{\rm{ren}}\left(p, e,\lambda,\mu\right)=0,
\end{equation}
where we have defined the $\beta$ and $\gamma$ functions as follows
\begin{equation}
\beta(e)= \lim_{\epsilon \rightarrow 0}\mu \frac{d e}{d \mu},\quad \beta(\lambda)= \lim_{\epsilon \rightarrow 0}\mu \frac{d \lambda}{d \mu},\quad   \gamma_{_{A}}(e)= \lim_{\epsilon \rightarrow 0}\frac{\mu}{2} \frac{d }{d \mu} \ln Z_{_{3}}.
\end{equation}
Moreover, if we can consider the other two two-point functions, we find the anomalous dimensions for the scalar and ghost fields, $\gamma_{_{\phi}}= \frac{\mu}{2} \frac{d }{d \mu} \ln Z_{_{2}}$ and $\gamma_c = \frac{\mu}{2} \frac{d }{d \mu} \ln  \widetilde{Z}_{_{3}}$, respectively.
We now proceed in computing the one-loop-order radiative corrections to the basic functions so that  the respective renormalization constants can be determined, allowing us to find the basic $\beta$ and $\gamma$ functions.


\section{Radiative corrections}
\label{sec:4}

In this section, we shall compute the one-loop correction to the scalar self-energy, polarization tensor, ghost self-energy, and three-point and four-point vertex parts $\left<\phi^\dag \phi A\right>$, $\left<\phi^\dag \phi A A\right>$, and $\left<\phi^\dag \phi \phi^\dag \phi\right>$. With these expressions, we shall then proceed to determine the $\beta$ function for the NC-SQED.
\subsection{One-loop scalar self-energy}
\begin{figure}[t]
\vspace{-1.8cm}
\includegraphics[width=10cm,height=4.5cm]{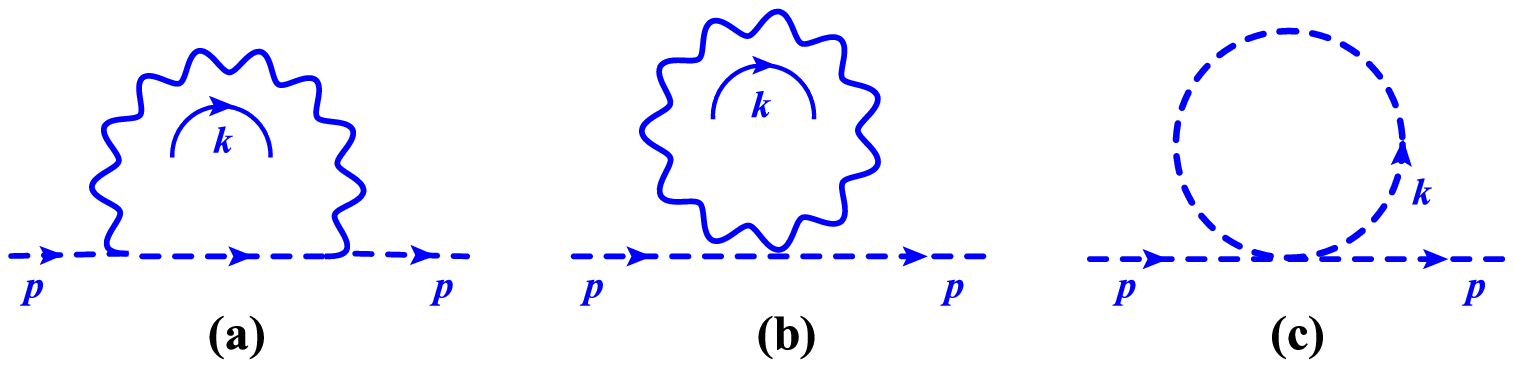}
 \centering \caption{One-loop self-energy graphs for the charged scalar field.}
\label{oneloopdiagrams1}
\end{figure}
We start by computing now the simplest one-loop contribution that is the scalar self-energy. The corresponding graphs
are depicted at Fig.~\ref{oneloopdiagrams1}. Their explicit expressions read
\begin{align}
\Sigma_{(a)}(p) & =-e^{2}\mu^{4-d}\int\frac{d^{d}k}{(2\pi)^{d}}\frac{(2p-k)^{2}}{k^{2}\left((p-k)^{2}-m^{2}\right)}, \nonumber \\
 \Sigma_{(b)}(p)& =2d e^{2}\mu^{4-d}\int\frac{d^{d}k}{(2\pi)^{d}}\frac{1}{k^{2}},\nonumber \\
\Sigma_{(c)} (p)&= \left(\lambda_1+\frac{\lambda_2}{2}\right)\mu^{4-d}\int\frac{d^{d}k}{(2\pi)^{d}} \frac{1 }{k^2-m^2}+ \frac{\lambda_2}{2}\mu^{4-d}\int\frac{d^{d}k}{(2\pi)^{d}}\frac{e^{ip\wedge k} }{k^2-m^2}
\end{align}
that the momentum integral on the contribution (b) is vanishing by dimensional regularization, i.e. $ \Sigma_{b}(p)=0$.
Moreover, we see that the contribution (a) is purely planar, while in the graph (c) noncommutative effects are present.
Besides, notice that the nonplanar term from $\Sigma_{(c)}$ is proportional to $\lambda_2 \left( |\tilde{p}|m\right)^{1-\frac{d}{2}} K_{1-\frac{d}{2}}\left( |\tilde{p}|m\right)$. But when the limit $d\rightarrow 4^+$ is taken, we see that this term is singular when $p\rightarrow 0$, i.e., $  \frac{1}{|\tilde{p}|^2}$ and $  \ln (|\tilde{p}|m)$, which is a manifestation of the known UV/IR mixing. Fortunately, this UV/IR mixing is fully removed in the case in which $\lambda_2=0$ \cite{Arefeva:2000uu}.

The remaining momentum integral can be computed by standard Feynman integration and dimensional regularization, and it results in
\begin{align}
\Sigma_{_{_{1-loop}}}(p)&=\Sigma_{(a)}(p)+\Sigma_{(b)}(p)+\Sigma_{(c)}(p) \nonumber \\
&=-\frac{ie^{2}}{4\pi^{2}\epsilon'}\left(p^{2}+\frac{m^{2}}{2}\right)+\frac{im^2}{8\pi^2\epsilon'} \left(\lambda_1+\frac{\lambda_2}{2}\right)+\rm{finite},
\end{align}
where we have defined by convenience $\frac{2}{\epsilon'}=\frac{2}{\epsilon}-\gamma+\log\left(4\pi\mu^{2}\right)$, and $\epsilon = 4-d \rightarrow 0^+$.
\noindent
Now, making use of the renormalized Lagrangian, we find that the renormalization conditions imply that
\begin{align}
-\frac{ie^{2}}{4\pi^{2}\epsilon'}\left(p^{2}+\frac{m^{2}}{2}\right)+\frac{im^2}{8\pi^2\epsilon'} \left(\lambda_1+\frac{\lambda_2}{2}\right)+i(\delta_{_{2}}p^{2}-\delta_{m})=0,
\end{align}
so the expressions for the counterterms are
\begin{align}
\delta_{_{2}}=\frac{e^{2}}{4\pi^{2}\epsilon'},\quad \delta_{m}=-\frac{e^{2}m^{2}}{8\pi^{2}\epsilon'}+\frac{m^2}{8\pi^2\epsilon'} \left(\lambda_1+\frac{\lambda_2}{2}\right).
\end{align}
As we observe, $Z_{2}$ does not receive any contribution from the self-interaction term for the scalar fields while the constant $Z_{m}$ is clearly changed.

In particular, we can then make use of $Z_{_{2}}=1+\delta_{_{2}}$ and find the renormalization constant of the matter sector as
\begin{align}
Z_{_{2}}=1+\frac{e^{2}}{4\pi^{2}\epsilon'},  \label{eq:0.6}
\end{align}
which is the same as in the commutative case.
\subsection{One-loop photon self-energy}
\begin{figure}[t]
\includegraphics[width=17cm,height=3.2cm]{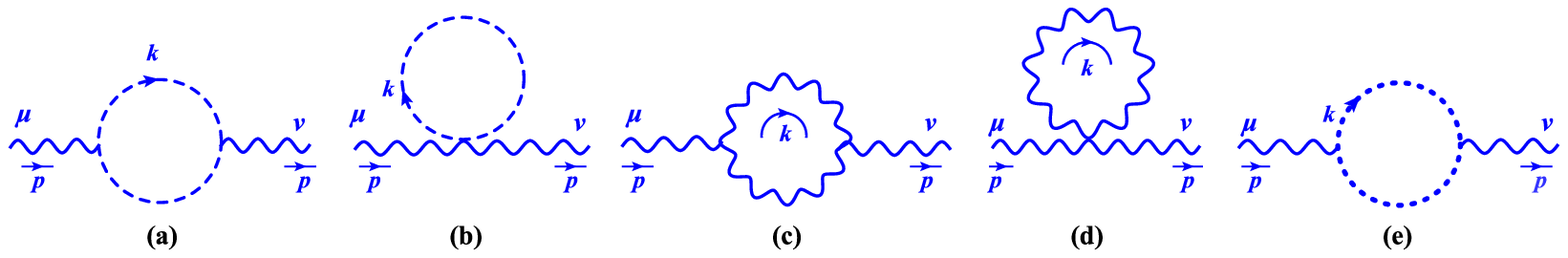}
 \centering\caption{One-loop self-energy graphs for the gauge field.}
\label{oneloopdiagrams2}
\end{figure}

We have five graphs contributing to the photon polarization tensor at one-loop order; these are given in Fig.~\ref{oneloopdiagrams2}.  The interaction among the gauge and charged scalar fields is given by graphs (a) and (b); graph (c) comes from the cubic self-coupling of the gauge-field; graph (d) is the tadpole contribution; and graph (e) comes from the ghost loop.
 Let us consider first the contributions from graphs (a) and (b):
\begin{align}
\Pi_{(a)}^{\mu\nu}(p)&=e^{2}\mu^{4-d}N_{B} \int\frac{d^{d}k}{(2\pi)^d}
\frac{(2k-p)^{\mu}(2k-p)^{\nu}}
{(k^{2}-m^{2})((k-p)^{2}-m^{2})}, \nonumber \\
\Pi_{(b)}^{\mu\nu}(p) & =-2e^{2}\mu^{4-d}N_{B}\int\frac{d^{d}k}{(2\pi)^{d}}\frac{g^{\mu\nu}}{k^{2}-m^{2}}.
\end{align}
Notice once again the absence of noncommutative effects in these contributions. It is convenient to rewrite these two contributions together so that using the Feynman parametrization method we have
\begin{align}
\Pi^{\mu\nu}_{a+b}(p)&=e^2\mu^{4-d}N_{B}
\int_0^{1}dy\int\frac{d^{d}Q}{(2\pi)^d}\bigg[\frac{(1+4y^2-4y)
p^{\mu}p^{\nu}+(-4y^2+6y-2)g^{\mu\nu}p^{2}}{(Q^2-\Delta)^2}\nonumber\\
&-\frac{2g^{\mu\nu}}{(Q^2-\Delta)}+\frac{\frac{4}{d}g^{\mu\nu}Q^{2}}{(Q^2-\Delta)^2}
\bigg],
\end{align}
where $Q=k-yp$ and $\Delta=y(y-1)p^{2}+m^{2}$.
The remaining integration can be readily evaluated and we find
\begin{align}
\Pi_{a+b}^{\mu\nu}(p)=\frac{ie^{2}}{24\pi^{2}\epsilon'}(p^{\mu}p^{\nu}-g^{\mu\nu}p^{2})N_{B}+\rm{finite},
\label{eq:a-b}
\end{align}
where $N_{_{B}}$ is the number of  independent scalar bosons with charge $\pm1$. As one should naively expect from the results from ordinary scalar QED, the
above result is consistent with the Ward identity.
The remaining three diagrams (c), (d), and (e) are given by
\begin{align}
\Pi_{(c)}^{\mu\nu} (p)& =e^{2}\mu^{4-d}C_{(c)}\int\frac{d^{d}k}{(2\pi)^{d}}\left(\frac{1-\cos{(k\wedge p)}}{k^{2}(p+k)^{2}}\right)N_{(c)}^{\mu\nu}, \\
\Pi_{(d)}^{\mu\nu}(p) & =e^{2}\mu^{4-d}C_{(d)}\int\frac{d^{d}k}{(2\pi)^{d}}\left(\frac{1-\cos{(k\wedge p)}}{k^{2}(p+k)^{2}}\right)N_{(d)}^{\mu\nu}, \\
\Pi_{(e)}^{\mu\nu}(p) & =e^{2}\mu^{4-d}C_{(e)}\int\frac{d^{d}k}{(2\pi)^{d}}\left(\frac{1-\cos\left({k\wedge p}\right)}{k^{2}(p+k)^{2}}\right)N_{(e)}^{\mu\nu},
\end{align}
where their tensor structures at the numerator are
\begin{align}
N_{(c)}^{\mu\nu} & =2\bigg[(d-6)p^{\mu}p^{\nu}+(2d-3)(p^{\mu}k^{\nu}+k^{\mu}p^{\nu})
+(4d-6)k^{\mu}k^{\nu}+(5p^{2}+2k^{2}+2p.k)g^{\mu\nu}\bigg],\nonumber \\
N_{(d)}^{\mu\nu} & =4g^{\mu\nu}(p+k)^{2}(d-1),\nonumber \\
N_{(e)}^{\mu\nu} & =2(p+k)^{\nu}k^{\mu},
\end{align}
and also notice that the symmetry factors of them are $C_{(c)}=\frac{1}{2}$, $C_{(d)}=\frac{1}{2}$ and $C_{(e)}=-1$, respectively.
Moreover, we can cast all three of these contributions into a common expression,
\begin{align}
\Pi_{c+d+e}^{\mu\nu}=\Pi_{(c)}^{\mu\nu}+\Pi_{(d)}^{\mu\nu}+\Pi_{(e)}^{\mu\nu},
\end{align}
so that it reads
\begin{align}
\Pi_{c+d+e}^{\mu\nu}=e^{2}\mu^{4-d}\int_{0}^{1}dx\int\frac{d^{d}Q}{(2\pi)^{d}}
\left(\frac{1-\cos{[Q\wedge p]}}{(Q^{2}-\Delta_1)^{2}}\right)N_{c+d+e}^{\mu\nu}, \label{eq:1.2}
\end{align}
the tensor structure is given as
\begin{align}
N_{c+d+e}^{\mu\nu} & =\bigg(d-6+x^{2}(4d-8)-x(4d-8)\bigg)p^{\mu}p^{\nu}+\bigg(3+2d+2d x^{2}-x(4d-2)\bigg)p^{2}g^{\mu\nu},\nonumber \\
 & +(4d-8)Q^{\mu}Q^{\nu}+2d Q^{2}g^{\mu\nu},
\end{align}
where $Q=k+xp$ and $\Delta_1=x(x-1)p^{2}$.

From Eq.~\eqref{eq:1.2}, we find for the first time the presence of noncommutative effects on the one-loop functions so that noncommutativity will engender new features on the $\beta$-function expression as we expected. We shall next compute separately the planar and nonplanar contributions. The planar part is evaluated straightforwardly, and it reads
\begin{align}
(\Pi_{c+d+e}^{\mu\nu})_{p}=\frac{ie^{2}}{8\pi^{2}\epsilon'}\frac{10}{3}
\bigg(p^{2}g^{\mu\nu}-p^{\mu}p^{\nu}\bigg)+\rm{finite}.  \label{eq:1.7}
\end{align}
The nonplanar contribution of Eq.~\eqref{eq:1.2} can be computed with the help of the integrals
\begin{equation}
\int\frac{d^{d}q}{\left(2\pi\right)^{d}}\frac{1}{\left(q^{2}-s^{2}\right)^{\alpha}}e^{ik\wedge q}=\frac{2i\left(-\right)^{\alpha}}{\left(4\pi\right)^{\frac{d}{2}}}\frac{1}{\Gamma\left(\alpha\right)}
\frac{1}{\left(s^{2}\right)^{\alpha-\frac{d}{2}}}\left(\frac{|\tilde{k}|s}
{2}\right)^{\alpha-\frac{d}{2}}K_{\alpha-\frac{d}{2}}\left(|\tilde{k}|s\right), \label{eq:C.1}
\end{equation}
and
\begin{align}
\int\frac{d^{d}q}{\left(2\pi\right)^{d}}\frac{q^{\mu}q^{\nu}}{\left(q^{2}-s^{2}\right)^{\alpha}}e^{ik\wedge q} & =\eta^{\mu\nu}F_{\alpha}+\frac{\tilde{k}^{\mu}\tilde{k}^{\nu}}{\tilde{k}^{2}}G_{\alpha},\label{eq:C.2}
\end{align}
where
\begin{align}
\left\{ F_{\alpha},G_{\alpha}\right\}  & =\frac{i\left(-\right)^{\alpha-1}}{\left(4\pi\right)^{\frac{d}{2}}}\frac{1}{\Gamma\left(\alpha\right)}
\frac{1}{\left(s^{2}\right)^{\alpha-1-\frac{d}{2}}}~\left\{ f_{\alpha},g_{\alpha}\right\},\label{eq:C.3}
\end{align}
with the definitions
\begin{align}
f_{a} & =\left(\frac{s|\tilde{k}|}{2}\right)^{\alpha-1-\frac{d}{2}}
K_{\alpha-1-\frac{d}{2}}\left(|\tilde{k}|s\right),\label{eq:C.3a}\\
g_{\alpha} & =\left(2\alpha-2-d\right)
\left(\frac{s|\tilde{k}|}{2}\right)^{\alpha-1-\frac{d}{2}}K_{\alpha-1-\frac{d}{2}}
\left(|\tilde{k}|s\right)-2\left(\frac{s|\tilde{k}|}{2}\right)^{\alpha
-\frac{d}{2}}K_{\alpha-\frac{d}{2}}\left(|\tilde{k}|s\right).\label{eq:C.3b}
\end{align}
Nonetheless, we find that the nonplanar part is UV finite when $\epsilon=4- \omega\rightarrow0^+$ and does not have any effect on the $\beta$ function. On the other hand, the nonplanar part is proportional to terms
$ e^2 \left( |\tilde{p}|m\right)^{1-\frac{d}{2}} K_{1-\frac{d}{2}}\left( |\tilde{p}|m\right)$ and $ e^2 \left( |\tilde{p}|m\right)^{2-\frac{d}{2}} K_{2-\frac{d}{2}}\left( |\tilde{p}|m\right)$, which in turn are singular when $d\rightarrow 4^+$ and $p\rightarrow 0$; i.e., they behave as $  \frac{1}{|\tilde{p}|^2}$ and $ \ln  (|\tilde{p}|m )$, which are a manifestation of the known UV/IR mixing.
Although the UV/IR mixing could be removed from the scalar sector by a choice of $\lambda_2=0$, we lack of a suitable set of parameters in order to remove it in the gauge field sector without making the theory trivial. Hence, in this sense, this UV/IR mixing jeopardize the dynamics of the gauge field.

It is notable that, although the nonplanar part does not contribute to the $\beta$ function, the planar part \eqref{eq:1.7} has an origin precisely from the noncommutativity, since graphs (c), (d), and (e) do not appear in the ordinary theory. In this way, noncommutativity effects are encoded in the contribution \eqref{eq:1.7}.

Hence, the complete one-loop-order correction to the photon self-energy is given by
\begin{align}
\Pi_{_{_{1-loop}}}^{\mu\nu}=\frac{ie^{2}}{16\pi^{2}\epsilon'}\left(\frac{20}{3}-\frac{2N_{B}}{3}\right)
\bigg(p^{2}g^{\mu\nu}-p^{\mu}p^{\nu}\bigg)+\rm{finite},
\end{align}
and by taking into account the renormalized polarization tensor, the renormalization condition shows that the counterterm $ \delta_{_{3}}$ satisfies the relation
\begin{align}
\frac{ie^{2}}{16\pi^{2}\epsilon}\left(\frac{20}{3}-\frac{2N_{B}}{3}\right)\bigg(p^{2}g^{\mu\nu}-p^{\mu}p^{\nu}\bigg)
-i\delta_{_{3}}(g^{\mu\nu}p^{2}-p^{\mu}p^{\nu})=0.
\end{align}
Finally, by means of $Z_{_{3}} = 1+\delta_{_{3}}$, we can compute the renormalization constant related to the photon sector
\begin{equation}
Z_{_{3}}=1+\frac{e^{2}}{16\pi^{2}\epsilon'}\left(\frac{20}{3}-\frac{2N_{B}}{3}\right).  \label{eq:1.6}
\end{equation}
This is one of the main results in order to determine the $\beta$ function.
However, one might wish to compare Eq.~\eqref{eq:1.6} with the expression from the commutative scalar QED, where the renormalization constant of the gauge field comes solely from Eq.~\eqref{eq:a-b} and is given by $Z_{_{3}}=1-\frac{e^{2}N_{B}}{24\pi^{2}\epsilon'}$.
It is worth mentioning that in commutative Yang-Mills gauge theories coupled to the scalar fields with an arbitrary number of bosons the renormalization constant of the non-Abelian gauge field is described by \cite{peskin}
\begin{equation}
Z_{_{3}}=1+\frac{g^{2}}{16\pi^{2}\epsilon'}\bigg(\frac{10}{3}C_{2}(G)-\frac{2N_{B}}{3}C(r)\bigg),
\end{equation}
where $C_{2}(G)$ is the quadratic Casimir operator of the adjoint representation and $C(r)$ is a constant for each representation.
\subsection{One-loop ghost self-energy}
\begin{figure}[t]
\vspace{-1.8cm}
\includegraphics[width=4.3cm,height=4.2cm]{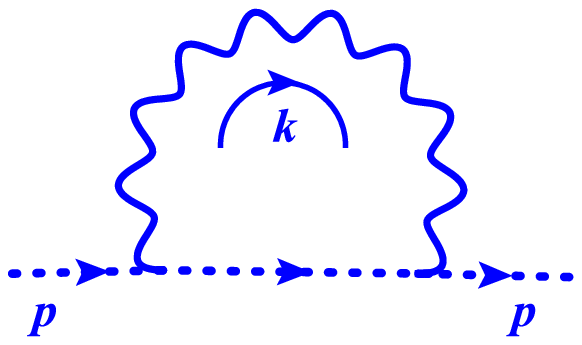}
 \centering\caption{One-loop self-energy graph for the ghost field.}
\label{ghost}
\end{figure}
 The calculation of the ghost self-energy is straightforward since it comes from only one graph, depicted in Fig.~\ref{ghost}.
 The given contribution reads
 \begin{equation}
 G\left(p\right) = 2e^{2}\mu^{4-d}\int\frac{d^{d}k}{(2\pi)^{d}} \frac{p.\left(p-k\right)}{k^2 \left(p-k\right)^2} \left(1-e^{ik\wedge p} \right).
 \end{equation}
 The planar part of this expression yields
 \begin{align}
 \left(G\right) _p\left(p\right) =\frac{ie^{2}}{8\pi^{2}\epsilon'}p^2+\rm{finite},
\end{align}
while the nonplanar part is UV finite. However, the nonplanar part is proportional to terms  $ e^2 \left( |\tilde{p}|m\right)^{2-\frac{d}{2}} K_{2-\frac{d}{2}}\left( |\tilde{p}|m\right)$, which in turn is singular when $d\rightarrow 4^+$ and $p\rightarrow 0$, i.e., behaving as $ \ln  (|\tilde{p}|m )$, which is a manifestation of the known UV/IR mixing.
Once again, this UV/IR mixing cannot be removed, although it does not jeopardize the dynamics of the nonphysical ghost field. Thus, the respective ghost renormalization constant by making use of $\widetilde{Z}_{_{3}}=1+\widetilde{\delta}_{_{3}}$ is given by
\begin{align}
\widetilde{Z}_{_{3}}=1+\frac{e^{2}}{8\pi^{2}\epsilon'},  \label{eq:1.10}
\end{align}
and it plays an important part in establishing the full set of renormalization constants.
\subsection{One-loop correction to the three-point vertex part $\left<\phi^\dag\phi A\right>$}
\begin{figure}[t]
\includegraphics[width=17cm,height=4.5cm]{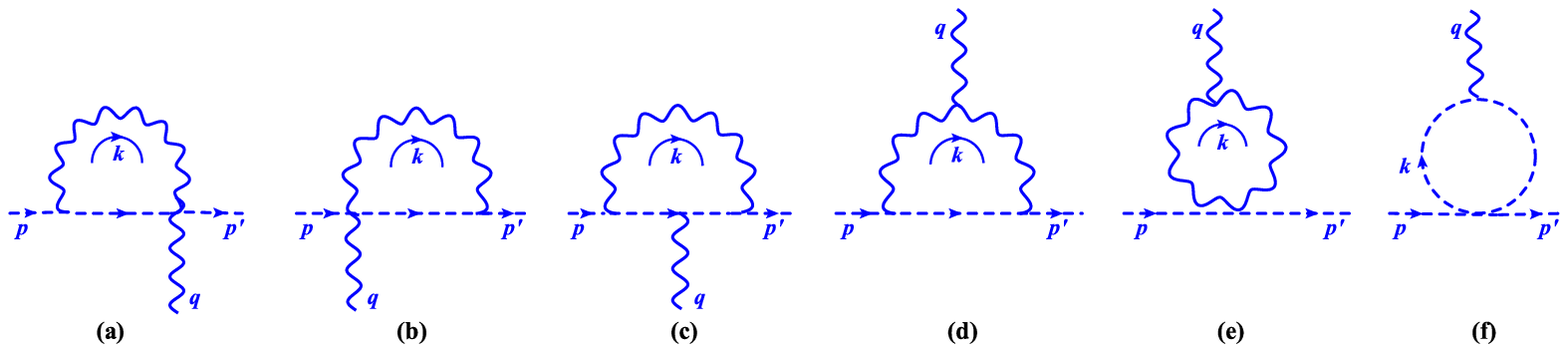}
 \centering  \caption{One-loop graphs contributing to three-point vertex $\left<\phi^\dag\phi A\right>$.}
\label{oneloopdiagrams3}
\end{figure}

We now turn our attention to the calculation of the one-loop correction to the
vertex part $\left<\phi^\dag\phi A\right>$. Here, $p$ and $p'$ are the momenta of the incident and emergent scalar fields,
and $q=p-p'$ is the transferred momentum to the gauge field. The corresponding diagrams are shown in Fig.~\ref{oneloopdiagrams3}.

Let us start by computing the contributions from graphs (a) and (b):
\begin{align}
\Lambda_{(a)}^{\mu}(p,p') & = e^{2}\mu^{4-d}\int\frac{d^{d}k}{(2\pi)^{d}}\frac{(2p-k)^{\mu}}
{k^{2}\left((p-k)^{2}-m^{2}\right)}e^{\frac{i}{2}p\wedge p'}\left(e^{ik\wedge q}+1\right), \label{eq:2.1}   \\
\Lambda_{(b)}^{\mu}(p,p')  &=e^{2}\mu^{4-d}\int\frac{d^{d}k}{(2\pi)^{d}}\frac{(2p'-k)^{\mu}}{k^{2}\left((p'-k)^{2}-m^{2}\right)} e^{\frac{i}{2}p\wedge p'}\left(e^{ik\wedge q}+1\right).
\end{align}
We observe that, apart from the phase factor $e^{\frac{i}{2}p\wedge p'}$, diagrams (a) and (b) are related by the symmetry replacement $p \leftrightarrow p'$. In this way we can compute the contribution of (b) based on the results of (a). We proceed now to compute separately the planar part of Eq.~\eqref{eq:2.1}
\begin{align}
  \Lambda_{(a)}^{\mu}\big\vert_{p}&=e^{2}\mu^{4-d}e^{\frac{i}{2}p\wedge p'}\int\frac{d^{d}k}{(2\pi)^{d}}\frac{(2p-k)^{\mu}}{k^{2}\left((p-k)^{2}-m^{2}\right)}\nonumber \\
&=\frac{3ie^{2}}{16\pi^{2}\epsilon'}~p^{\mu}e^{\frac{i}{2}p\wedge p'}+\rm{finite}, \label{eq:2.2}
\end{align}
 and with help of Eq.~\eqref{eq:C.1}, we can compute the nonplanar part
\begin{align}
\Lambda_{(a)}^{\mu}\big\vert_{n-p}=e^{2}\mu^{4-d}e^{\frac{i}{2}p\wedge p'}\int\frac{d^{d}k}{(2\pi)^{d}}\frac{(2p-k)^{\mu}}{k^{2}\left((p-k)^{2}-m^{2}\right)}
e^{ik\wedge q},
\end{align}
and show that it is also UV finite when $\epsilon=4-d\rightarrow0^+$.
Based on the above discussion, on the symmetry argument, we can determine the contribution of graph (b) from Eq.~\eqref{eq:2.2} as being
\begin{align}
\Lambda_{(b)}^{\mu}\big\vert_{p}&=\frac{3ie^{2}}{16\pi^{2}\epsilon'}~p'^{\mu}e^{\frac{i}{2}p\wedge p'}+\rm{finite},\\
 \Lambda_{(b)}^{\mu}\big\vert_{n-p}&=\rm{finite}.
\end{align}
Finally, we find that the contributions of the diagrams (a) and (b) can be expressed conveniently as
\begin{align}
\Lambda_{(a+b)}^{\mu}\big\vert_{p} & =\frac{3ie^{2}}{16\pi^{2}\epsilon'}(p+p')^{\mu}e^{\frac{i}{2}p\wedge p'}+\rm{finite}.
\label{eq:2.3}
\end{align}

Moreover, the expression of the graph (c) is given by
\begin{align}
\Lambda_{(c)}^{\mu}(p,p')=-e^{2}\mu^{4-d}e^{\frac{i}{2}p\wedge p'}\int\frac{d^{d}k}{(2\pi)^{d}}\frac{(2p-k).(2p'-k)(p+p'-2k)^{\mu}}{k^{2}((p'-k)^{2}-m^{2})((p-k)^{2}-m^{2})} e^{ik\wedge q},
\end{align}
and we can perform the momentum integration with help of Eqs.~\eqref{eq:C.1} and \eqref{eq:C.2}. But first notice that this graph is purely nonplanar
and UV finite when the limit of $\epsilon=4-d\rightarrow0^+$ is taken. Hence, this diagram does not contribute to the $\beta$ function
\begin{align}
\Lambda_{(c)}^{\mu}(p,p')=\rm{finite}, \label{eq:2.4}
\end{align}
while its  commutative counterpart is divergent in this limit.
So far, we have computed diagrams similar to those of ordinary scalar QED. The following contributions, graphs (d) and (e), are engendered by noncommutativity, so we shall actually obtain the noncommutative contribution to this vertex function.
 We compute next the diagram (d), the expression of which is
\begin{align}
\Lambda^{\mu}_{(d)}=e^{2}\mu^{4-d}e^{\frac{i}{2}(p\wedge
p')}\Gamma(3)\int_0^{1}dy\int_0^{1-y}dz\int\frac{d^{d} Q}{(2\pi)^d}\frac{\left(1-e^{-iq\wedge
(Q+zp)}\right)N^{\mu}_{(d)}}{(Q^2-\Delta)^{3}},
\end{align}
where $\Delta=(yq+zp)^2-yq^2-zp^2+zm^2$, $Q=k-yq-zp$, and the numerator $N^{\mu}_{(d)}$ is defined as
\begin{align}
N^{\mu}_{(d)}&=\big[Q-(1+y)p'+(1+y+z)p\big].\big[(1-y-z)p+(1+y)p'-Q\big]\big[(2-z-y)p+yp'-Q\big]^{\mu}\nonumber\\
&+\big[(2-z-y)p+yp'-Q\big].\big[Q+(2-y)p'+(z+y-2)p\big]\big[(1-y-z)p+(1+y)p'-Q\big]^{\mu} \nonumber\\
&+\big[(2-z-y)p-Q+yp'\big].\big[(1-y-z)p+(1+y)p'-Q\big]\big[(1-2y-2z)p-(1-2y)p'-2Q\big]^{\mu}.
\end{align}
We can proceed and separate the different powers of $Q$ in the numerator so that after computing the momentum and Feynman integrals we obtain the result for the planar part,
\begin{align}
\Lambda_{(d)}^{\mu}\big\vert_{p}=-\frac{3ie^{2}}{16\pi^{2}\epsilon'}
(p+p')^{\mu}e^{\frac{i}{2}p\wedge p'} +\rm{finite},
\end{align}
while we see that the nonplanar contribution is UV finite in this case $ \Lambda_{(d)}^{\mu}\big\vert_{n-p}=\rm{finite}$. Hence, the contribution of this diagram to the respective renormalization constant is precisely
\begin{align}
\Lambda_{(d)}^{\mu}=-\frac{3ie^{2}}{16\pi^{2}\epsilon'}
(p+p')^{\mu}e^{\frac{i}{2}p\wedge p'} +\rm{finite}. \label{eq:2.5}
\end{align}
This is the first noncommutative contribution to this vertex function (also to the respective renormalization constant). We shall next compute the diagram (e), which is also due to noncommutativity. The graph (e) has the expression
\begin{align}
\Lambda_{(e)}^{\mu}=-2e^{2}\mu^{4-d} (1-d)e^{\frac{i}{2}p\wedge p'}\int_{0}^{1}dy\int\frac{d^{\omega}Q}{(2\pi)^{d}}~\frac{2Q^{\mu}e^{iq\wedge Q}}{(Q^{2}-\Delta)^{2}},
\end{align}
with $Q=k-yq$ and $\Delta=y(y-1)q^{2}$. Once again, we see a purely nonplanar graph, so when the limit $\epsilon=4-d\rightarrow0^+$ is taken, we get a UV finite result
\begin{align}
\Lambda_{(e)}^{\mu} =\rm{finite}. \label{eq:2.6}
\end{align}
Finally, we consider graph (f),
\begin{align}
\Lambda^{\mu}_{(f)}&=e \int\frac{d^{d}k}{(2\pi)^{d}}
\frac{(q-2k)^{\mu}}{(k^2-m^2)((q-k)^2-m^2)}e^{\frac{i}{2}q\wedge
k}\nonumber\\
&\times\bigg[ \lambda_1 \cos{\big(\frac{p'\wedge
 p+q\wedge k}{2}\big)}+\lambda_2 \cos{\big(\frac{p'\wedge
k}{2}\big)}\cos{\big(\frac{p\wedge q-p\wedge k}{2}\big)}\bigg ]
\end{align}
the contribution of which is UV finite when the limit $\epsilon=4-d\rightarrow0^+$ is taken, so that we have
\begin{align}
\Lambda_{(f)}^{\mu} =\rm{finite}. \label{eq:2.6}
\end{align}
Notice that the finiteness of this result is independent of the couplings $e$, $\lambda_1$, and $\lambda_2$.

Therefore, the total one-loop contribution is found by summing all the obtained results Eqs.~\eqref{eq:2.3}, \eqref{eq:2.4}, \eqref{eq:2.5}, and \eqref{eq:2.6}; thus,
\begin{align}
\Lambda_{_{_{1-loop}}}^{\mu}(p)=\rm{finite}.
\end{align}
 We note that the divergent parts of Eqs.~\eqref{eq:2.3} and \eqref{eq:2.5} are mutually cancelled, so the one-loop-order correction to the vertex part $\Lambda_{_{1-loop}}^{\mu}(p)$, corresponding to the vertex $\left<\phi^\dag\phi A\right>$, is finite due to noncommutative effects. This result is in agreement with the analysis performed in Ref.~\cite{Arefeva:2000uu}, in which the authors have shown that noncommutative scalar QED$_4$ in adjoint representation is one-loop renormalizable.
  This result is in contrast to its commutative counterpart which gives a divergent contribution to the vertex part.
In fact, since we have a UV finite expression for the one-loop correction, the renormalization condition for the vertex part yields
\begin{equation}
\Lambda_{_{1-loop}}^{\mu}(p)\bigg\vert_{_{div.}}-i\delta_{_{1}}(p+p')^{\mu}e^{\frac{i}{2}p\wedge p'}=0,
\end{equation}
thus we conclude that $\delta_{_{1}}=0$, and consequently the renormalization constant reads
\begin{equation}
Z_{_{1}}=1. \label{eq:2.7}
\end{equation}
Surprisingly we see that the noncommutative contribution to the vertex part $\Lambda_{_{_{1-loop}}}^{\mu}(p)$ is such that the ordinary divergent terms are exactly cancelled, rendered a UV finite expression at one-loop order, unlike the commutative result which reads $Z_{_{1}}\vert_{\rm com.}=1+\frac{e^{2}}{4\pi^{2}\epsilon'}$.

\subsection{One-loop correction to the four-point vertex part $\left<\phi^\dag\phi A A\right>$}
\begin{figure}[t]
\includegraphics{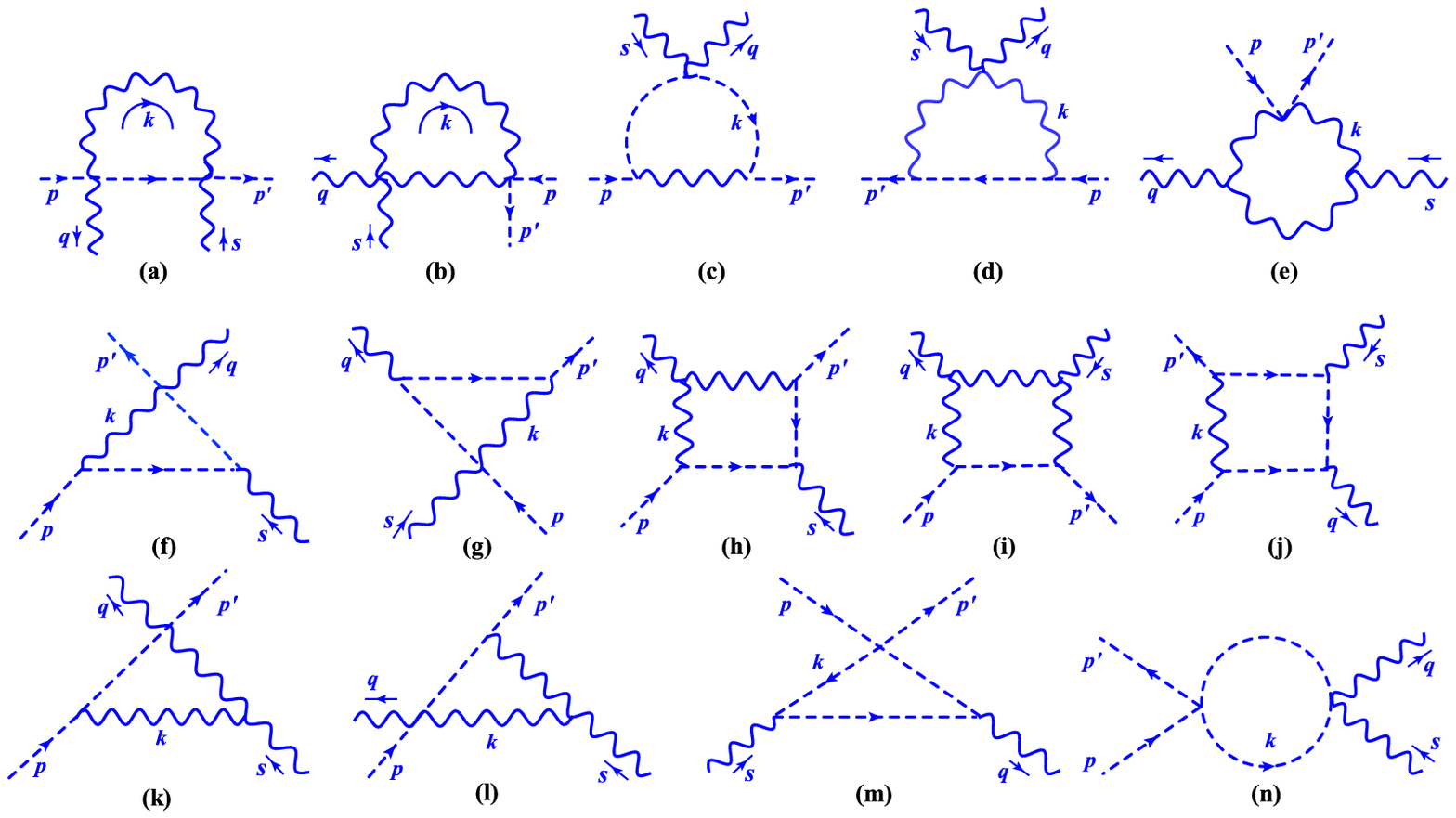}
 \centering \caption{One-loop graphs contributing to four-point vertex $\left<\phi^\dag\phi A A\right>$.}
\label{oneloopdiagrams4}
\end{figure}

The last point that we will discuss in this section, in order to compute the gauge coupling $\beta$ function, is the
calculation of the  four-point vertex part $\left<\phi^\dag\phi A A\right>$, which is the fourth basic function and is
related to Compton's scattering. The 12 diagrams that
contribute to this function at one-loop order are shown in the
Fig.~\ref{oneloopdiagrams4}.
We realize that the diagrams (a), (c), (f), (g), (j), (m), and (n) are the same from the ordinary commutative theory, while
the diagrams (b), (d), (e), (h), (i), (k), and (l) originated from the noncommutativity. For this reason, we will discuss these two classes of diagrams separately in order to highlight the part played by the noncommutative effects.

Let us summarize the expressions corresponding to the (commutative) diagrams (a), (c), (f), (g), (j), (m), and (n) written in a simplified form:
 \begin{align}
\Xi^{\mu\nu}_{(a)}&=-e^2 \mu^{4-d}g^{\mu\nu}\ e^{\frac{i}{2}p\wedge p'} e^{\frac{i}{2}s\wedge q} \int\frac{d^{d}k}{(2\pi)^{\omega}}\frac{ \left[1+ e^{ik\wedge q}+ e^{ik\wedge (q-s)}+e^{-ik\wedge s}\right]}{k^2\left((p'-s-k)^2-m^2\right)},  \label{eq:4.1}\\
\Xi^{\mu\nu}_{(c)}&=e^2\mu^{4-d}g^{\mu\nu}  e^{-\frac{i}{2}p'\wedge p}\left(e^{\frac{i}{2}s\wedge q}+e^{-\frac{i}{2}s\wedge q}\right)  \nonumber \\
&\times  \int\frac{d^{d}k}{(2\pi)^{d}}\frac{( k+q-p'-s).(k-p)}{\left((p+k\right)^2\left(k^2-m^2\right)\left(\left(s-q-k\right)^2-m^2\right)} e^{\frac{i}{2}k\wedge (p' -2p)}, \label{eq:4.2}\\
\Xi^{\mu\nu}_{(f)}&= e^2\mu^{4-d}e^{\frac{i}{2}q\wedge
p'}e^{\frac{i}{2}p\wedge s} \nonumber \\
&\times \int\frac{d^{d}k}{(2\pi)^{d}}
\frac{ \left(2p-k\right)^{\mu}\left(2p-2k+s\right)^{\nu} }{k^2\left(\left(p-k+s\right)^2-m^2\right)\left(\left(p-k\right)^2-m^2\right)} \big[e^{i(s-q)\wedge k}+e^{is\wedge k}\big], \label{eq:4.3}
 \end{align}
 \begin{align}
\Xi^{\mu\nu}_{(g)}&  =e^2\mu^{4-d} e^{\frac{i}{2}q\wedge p'}e^{\frac{i}{2}p\wedge s} \nonumber \\
&\times \int\frac{d^{d}k}{(2\pi)^{d}}
\frac{ \left(2p'-k\right)^{\mu}\left(p'+p-2k+s\right)^{\nu} }{k^2\left(\left(p-k+s\right)^2-m^2\right)\left(\left(p'-k\right)^2-m^2\right)}
 \big[e^{i(s-q)\wedge k}+e^{iq\wedge k}\big],\label{eq:4.4}\\
 \Xi^{\mu\nu}_{(j)}& =-e^2\mu^{4-d} e^{\frac{i}{2}p\wedge s}
e^{\frac{i}{2}q \wedge p'}  \nonumber \\
& \times \int\frac{d^{d}k}{(2\pi)^{d}} \frac{\left(2p'-2k+q\right)^{\mu}  \left(2p-2k+s\right)^{\nu}\left(2p-k\right).\left(2p'-k\right)} {k^2\left(\left(p-k\right)^2-m^2\right)\left(\left(p-k+s\right)^2-m^2\right)\left(\left(p'-k\right)^2-m^2\right)} e^{i(s-q )\wedge k},  \label{eq:4.5}
\end{align}
\begin{align}
\Xi^{\mu\nu}_{(m)}&=e^2\int\frac{d^{d}k}{(2\pi)^{d}}
\frac{(q-2k)^{\mu}(2q-s-2k)^{\nu}}{((q-k)^2-m^2)((q-s-k)^2-m^2)(k^2-m^2)}
e^{-\frac{i}{2}q\wedge k}e^{\frac{i}{2}(-s)\wedge (q-k)} \nonumber\\
&\times  \Big[ \lambda_1 \cos{\big(\frac{p'\wedge p+k\wedge(p'-p)}{2}\big)}+\lambda_2
\cos{\big(\frac{p'\wedge k}{2}\big)}\cos{\big(\frac{p\wedge
p'-k\wedge p}{2}\big)} \Big] \label{eq:4.18}\\
\Xi^{\mu\nu}_{(n)}&=2ie^2g^{\mu\nu}\int\frac{d^{d}k}{(2\pi)^{d}}
\frac{1}{(k^2-m^2)((p-p'+k)^2-m^2)}e^{\frac{i}{2}k\wedge
(p-p')} \cos{\big(\frac{q \wedge s}{2}\big)}  \nonumber\\
&\times   \Big( \lambda_1\cos{\big(\frac{p'\wedge p+(p-p'+k)\wedge
k}{2}\big)} +\lambda_2\cos{\big(\frac{p' \wedge
(p-p'+k)}{2}\big)}\cos{\big(\frac{p \wedge k}{2}\big)}\Big). \label{eq:4.19}
\end{align}

By a simple analysis, we readily conclude that the contributions from diagrams (c), (f), (g) and (j), Eqs.~\eqref{eq:4.2}, \eqref{eq:4.3}, \eqref{eq:4.4}, and \eqref{eq:4.5} are nonplanar, which actually means that they are UV  finite when $\epsilon = 4-d \rightarrow0^+$, showing that they do not contribute to the renormalization constant $Z_{_{4}}$ (i.e., to the $\beta$ function).
On the other hand, graphs (m) and (n), Eqs.~\eqref{eq:4.18} and \eqref{eq:4.19} have both planar and nonplanar parts. As usual, the nonplanar parts are UV finite when $\epsilon = 4-d \rightarrow0^+$, showing again that they do not contribute to the $\beta$ function. Now, if we compute the planar parts of both graphs, we can show that, though they are separately UV divergent when $\epsilon = 4-d \rightarrow0^+$, their sum is actually finite
\begin{align}
\Xi^{\mu\nu}_{(m)}+ \Xi^{\mu\nu}_{(n)}= \rm{finite}. \label{eq:4.20}
\end{align}
At last, we consider the remaining contributions. For graph (a), Eq.~\eqref{eq:4.1}, we find a contribution from the planar part to the $\beta$ function, so the full expression reads
\begin{align}
\Xi^{\mu\nu}_{(a)}(p)&= -\frac{ie^{2} }{4\pi^2\epsilon'}
g^{\mu\nu}e^{\frac{i}{2}p\wedge p'} \cos{\left(\frac{q\wedge s}{2}\right)}+\rm{finite}. \label{eq:4.6}
\end{align}

The remaining contributions, which consist of the second class of diagrams (b), (e), (i), (k) and (l), coming from the noncommutativity, are given as
\begin{align}
\Xi^{\mu\nu}_{(b)}& =\frac{3ie^2 }{4\pi^2 \epsilon'} g^{\mu\nu}e^{\frac{i}{2}p\wedge
p'}\cos{\big(\frac{q\wedge s}{2}\big)}+\rm{finite}, \label{eq:4.7} \\
\Xi^{\mu\nu}_{(e)}&=\frac{9ie^2 }{16\pi^2\epsilon'}g^{\mu\nu}e^{\frac{i}{2}p\wedge
p'}\cos{\big(\frac{q\wedge s}{2}\big)}+\rm{finite},  \label{eq:4.8}\\
\Xi^{\mu\nu}_{(i)} &=\frac{-3ie^2 }{16\pi ^2\epsilon'}g^{\mu\nu}
e^{\frac{i}{2}p\wedge p'}\cos{\big(\frac{q\wedge s}{2}\big)}+\rm{finite}, \label{eq:4.9}\\
\Xi^{\mu\rho}_{(k)}&= \frac{-3ie^2}{16\pi^2\epsilon}g^{\mu\nu}
e^{\frac{i}{2}p\wedge p'}\cos{\big(\frac{q\wedge s}{2}\big)} +\rm{finite},  \label{eq:4.10}\\
\Xi^{\mu\nu}_{(\ell)}&= \frac{-3ie^2 }{16\pi^2\epsilon}g^{\mu\nu}
e^{\frac{i}{2}p\wedge p'}\cos{\big(\frac{q\wedge s}{2}\big)} +\rm{finite},  \label{eq:4.11}
\end{align}
while the results from graphs (d) and (h) are purely nonplanar, and thus give simply a finite contribution
\begin{align}
\Xi^{\mu\nu}_{(d)} &= \rm{finite}, \label{eq:4.12} \\
 \Xi^{\mu\nu}_{(h)} &= \rm{finite}.   \label{eq:4.13}
\end{align}
We can finally determine the complete one-loop-order correction to the four-point vertex function by summing all 12 contributions computed above so that it yields
 \begin{align}
\Xi^{\mu\nu}_{_{_{1-loop}}}&=\frac{ie^2}{2\pi^2 \epsilon'}g^{\mu\nu} e^{\frac{i}{2}p\wedge
p'}\cos{\left(\frac{s\wedge q}{2}\right)}+\rm{finite}.
\end{align}
The respective renormalization condition for the four-point vertex function implies that the divergent part satisfies the relation
\begin{align}
\left.\Xi^{\mu\nu}_{_{_{1-loop}}}\right|_{_{div.}} +2ie^2\delta_{_{4}}g^{\mu\nu}
e^{\frac{i}{2}p\wedge p'}\cos{\left(\frac{s\wedge q}{2}\right)}=0,
\end{align}
so that the counterterm $\delta_{_{4}}$ can readily be evaluated.

Hence, the relevant renormalization constant is obtained by $ Z_{_{4}}=1+\delta_{_{4}}$,
\begin{align}
Z_{_{4}}=1-\frac{e^2}{4\pi^2 \epsilon'}.  \label{eq:4.17}
\end{align}

Although the quartic vertex $\left<\phi^\dag\phi \phi^\dag\phi\right>$ has played an important part when investigating the IR behavior of the scalar field self-energy, it has no effect whatsoever in the remaining terms, i.e., in the physical quantities involving the $\beta$ and $\gamma$ functions. We shall now consider the correction to the vertex function and highlight its fundamental role with regard to renormalizability.
\subsection{One-loop correction to the four-point vertex part $\left<\phi^\dag\phi \phi^\dag\phi\right>$}
\begin{figure}[t]
\includegraphics[width=10cm,height=8cm]{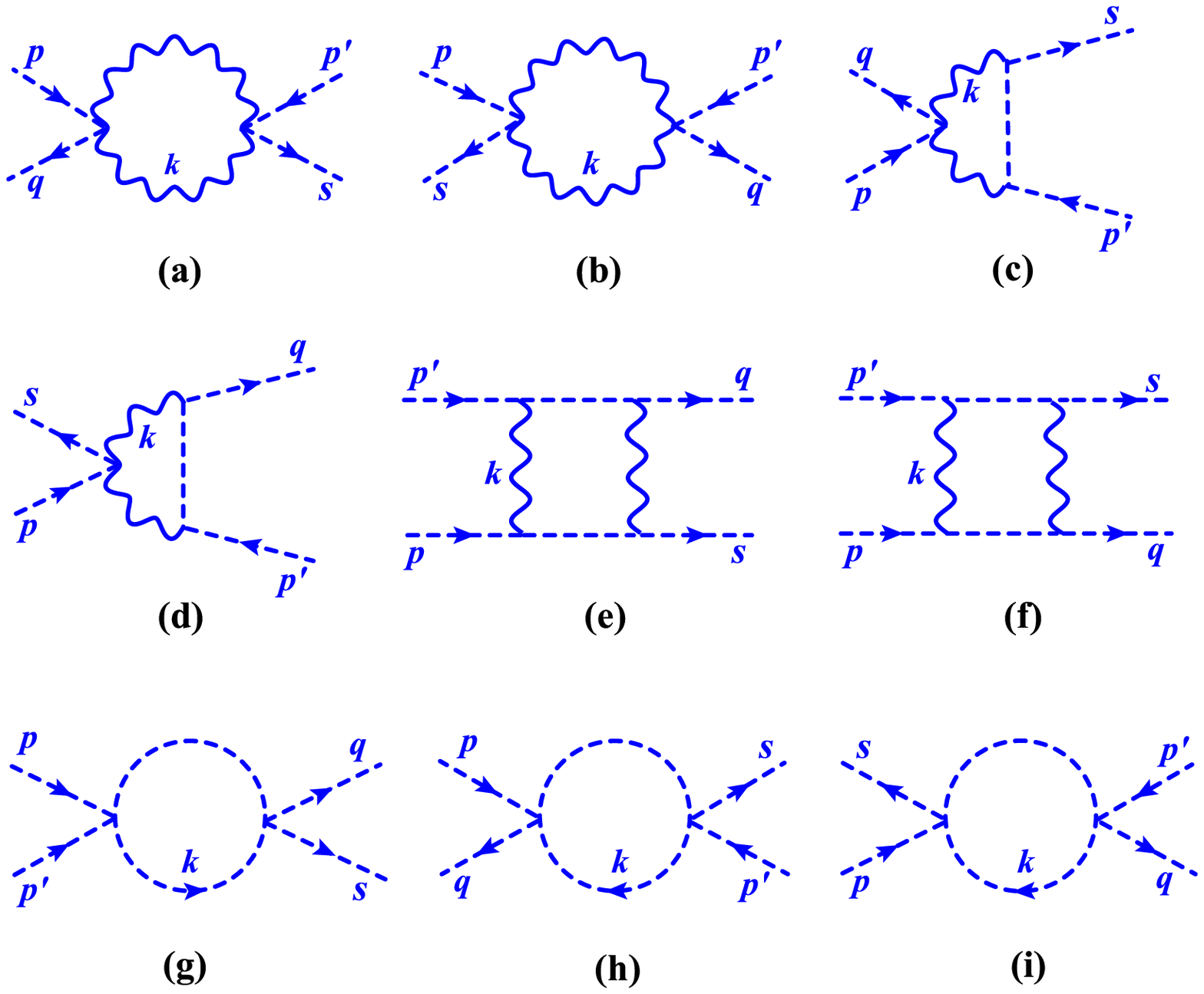}
 \centering \caption{One-loop graphs contributing to four-point vertex $\left<\phi\phi^\dag\phi\phi^\dag\right>$.}
\label{oneloopdiagrams5}
\end{figure}
Although we have computed all the relevant renormalization constants in order to compute the gauge coupling $\beta$ function, we now proceed in computing the renormalization constant related to the $\lambda_1$ and $\lambda_2$ couplings. For that matter, we consider the one-loop graphs (a)--(i) depicted in Fig.~\ref{oneloopdiagrams5}.

Let us start with the simplest contribution, which actually come from the gauge coupling, graphs (a)--(d). We see that the graphs (a) and (b) are related due to crossing symmetry, and the same is true for (c) and (d). Hence, we can compute the contributions (a) and (c) as
\begin{align}
\Upsilon_{(a)}&= 2e^4e^{\frac{i}{2}p \wedge
q}e^{\frac{i}{2}p' \wedge s} \int\frac{d^{d}k}{(2\pi)^{d}}
\frac{d}{(p-q-k)^2k^2}  \left( 1+\cos{\Big(k\wedge
(q-p)\Big) }\right) \\
\Upsilon_{(c)}&=-e^4e^{\frac{i}{2}p \wedge q}e^{\frac{i}{2}p' \wedge s} \int\frac{d^{d}k}{(2\pi)^{d}}\frac{(p'+s-k).(2s-k)}{k^2((s-k)^2-m^2)(p-q-k)^2}
  \Big( 1+ e^{i(p-q) \wedge k}  \Big)
\end{align}
It is easy to notice that the nonplanar parts of the above contributions are UV finite and hence do not contribute to the renormalization constant. Hence, focusing in the divergent part of the planar part, we find that the contribution from the (a), (b), (c), and (d) graphs is
\begin{align}
\Upsilon_{\rm{ (a+b+c+d)}}=\frac{3ie^4}{4\pi^2\epsilon'}\cos{\Big(\frac{p\wedge
q+p'\wedge s}{2}\Big)}+\rm{finite} \label{lambda1}
\end{align}
Moreover, the contributions (e) and (f) are nonplanar (due to noncommutativity), so they are UV finite, not contributing to the renormalization constant. At last, the contributions (g), (h), and (i) are more complex due to the vertex structure but are not related to each other. Computing the divergent part of the planar part of these contributions, we find
\begin{align}
\Upsilon _{\rm{ (g+h+i)}}=\frac{i}{8\pi^2\epsilon '}\Big[\left(\lambda_1^2+\frac{\lambda_2^2}{4}\right)\cos{\Big(\frac{q\wedge
p+s\wedge p'}{2}\Big)}+ \left(\lambda_1\lambda_2+\frac{\lambda_2^2}{4}\right)\cos{\big(\frac{p'\wedge
p}{2}\big)}\cos{\Big(\frac{s\wedge q}{2}\Big)} \Big] \label{lambda2}
\end{align}
Hence, summing up all the contributions, Eqs.\eqref{lambda1} and \eqref{lambda2}, we finally find the total contribution to the vertex function
\begin{align}
\Upsilon _{\rm{total}}&=\frac{i}{8\pi^2\epsilon'}\Big[\left(\lambda_1^2+\frac{\lambda_2^2}{4}+6e^4\right)\cos{\Big(\frac{q\wedge
p+s\wedge p'}{2}\Big)} \nonumber \\
&+ \left(\lambda_1\lambda_2+\frac{\lambda_2^2}{4}\right)\cos{\big(\frac{p'\wedge
p}{2}\big)}\cos{\Big(\frac{s\wedge q}{2}\Big)} \Big]. \label{lambda3}
\end{align}
The renormalization constants are then found by applying the respective conditions
\begin{align}
 Z_{\lambda_1} =1+\frac{1}{8\pi^2\epsilon'}  \left[\lambda_1+\frac{\lambda_2^2}{4\lambda_1}+\frac{6e^4}{\lambda_1}\right], \quad  Z_{\lambda_2} =1+\frac{1}{8\pi^2\epsilon'}  \left[\lambda_1+\frac{\lambda_2 }{4 } \right] \label{lambda4}
\end{align}
The divergent structure of these renormalization constants corroborates the need to add the quartic self-coupling for the charged fields, the primitive divergent vertices of which yield the renormalization of such coupling constants.

With this expression we conclude the section regarding the calculation of radiative correction, and next we use the renormalizability analysis previously established in order to compute basic $\beta$ and $\gamma$ functions.

\section{$\beta$ function and anomalous dimensions}
\label{sec:5}

Finally, based on the renormalization analysis developed in Sec.~\ref{sec:3}, we have that the $\beta$ function in this case is given by
\begin{equation}
\beta(e)=\mu\frac{d e(\mu)}{d \mu} , \quad \beta(\lambda)=\mu\frac{d \lambda(\mu)}{d \mu} ,
\end{equation}
where we recall that the relation between bare and dressed couplings is given either by $e_{0} =eZ_{_{3}}^{-1/2}Z_{_{2}}^{-1}Z_{_{1}}$ or alternatively by $e Z_{_{4}}^{1/2}Z_{_{3}}^{-1/2}Z_{_{2}}^{-1/2}$.

Hence, by making use of the results for the renormalization constants $Z_{_{2}}$, $Z_{_{3}}$, $\widetilde{Z}_{_{3}}$, $Z_{_{1}}$ and $Z_{_{4}}$, Eqs.~\eqref{eq:0.6}, \eqref{eq:1.6}, \eqref{eq:1.10}, \eqref{eq:2.7} and \eqref{eq:4.17}, respectively, we find for the one-loop order $\beta(e)$
\begin{align}
\beta(e)\bigg\vert_{\rm{NC-SQED}}=-\frac{e^{3}}{16\pi^{2}}\left(\frac{22}{3}-\frac{N_{B}}{3}\right),
\label{beta-14}
\end{align}
and from the expressions for $ Z_{\lambda_1} $ and $ Z_{\lambda_2} $  Eqs.\eqref{lambda4} we obtain the one-loop $\beta$ function for couplings of the scalar self-interaction sector
\begin{align}
\beta(\lambda_1) = \frac{1}{8\pi^2 }  \left[\lambda_1^2+\frac{\lambda_2^2}{4 }+6e^4- 4e^2\lambda_1\right], \quad
 \beta(\lambda_2) = \frac{1}{8\pi^2 }  \left[\lambda_1\lambda_2 +\frac{\lambda_2^2}{4 }-  4e^2\lambda_2\right].
\label{beta-15}
\end{align}
We observe that for small values of $N_{B}$ in Eq.~\eqref{beta-14} the sign of $\beta(e)$ is negative, and consequently the theory is asymptotically free, similar to the $\beta$ function of the non-Abelian gauge theories coupled to the matter with a small number of boson or fermion flavors. Furthermore, the result \eqref{beta-15} shows that the sign of $\beta(\lambda_1)$ and $\beta(\lambda_2)$ in the absence of the gauge fields is positive, as we expected \cite{weinberg-2}.
Moreover, the anomalous dimensions of the scalar, gauge, and ghosts fields are readily obtained
\begin{equation}
\gamma_{_{\phi}}=\frac{e^2}{8\pi^2},\quad \gamma_{_{A}}=\frac{e^2}{16\pi^2} \left(\frac{10}{3}-\frac{N_B}{3}\right),\quad \gamma_c=\frac{e^2}{32\pi^2}.
\end{equation}

The result \eqref{beta-14} can be compared to the one-loop $\beta(e)$ of NC-QED
\cite{hayakawa-1},
\begin{align}
\beta(e)\bigg\vert_{\rm{NC-QED}} = -\frac{e^{3}}{16\pi^{2}}\left(\frac{22}{3}-\frac{4N_{F}}{3}\right).
\end{align}
We see that both results are similar and just the contribution of the
matter sector is different. Indeed, in the absence of the matter sector, the contribution
of the gauge part to the $\beta(e)$ is the same, and this shows
the correctness of our result.
Besides, the coefficient appearing in the matter part of the NC-QED
is four times that of the NC-SQED. This, indeed, comes from the
trace over the gamma matrices in $d=4$, which indicates the spinor nature of the matter
field in the NC-QED that actually, when compared to the NC-SQED, has intrinsic angular
momentum (spin).

We may as well compare the present result \eqref{beta-14} with the $\beta$ function
of the commutative Yang-Mills theory coupled to the scalar fields
\begin{align}
\beta(e)\bigg\vert_{{\rm{SYM}}}= -\frac{e^{3}}{16\pi^{2}}\left(\frac{11}{3}C_{2}\left(G\right)-
\frac{ 2N_{B}}{3}C\left(r\right)\right),
\label{yang-mills}
\end{align}
in which for gauge fields in the adjoint representation $C_{2}( G ) = N_{c}$, while for matter fields in the fundamental (or antifundamental) representation, $C(r)=\frac{1}{2}$. As we see, the comparison of the results \eqref{beta-14} and \eqref{yang-mills} yields $C_{2}( G )=2$.
Consequently, the obtained result for the $\beta(e)$ of the NC-SQED \eqref{beta-14} is exactly the same as the $\beta$ function for the Yang-Mills theory in the presence of the scalar matter fields with the $SU(2)$ gauge group \cite{peskin,Capri:2015pxa}.
\section{Concluding remarks}
\label{sec:6}

In this paper, we have considered the scalar QED defined in a noncommutative spacetime, established its renormalizability, and computed the one-loop-order radiative corrections.
In addition, we have shown the BRST Slavnov transformations leaving the gauge-fixed Lagrangian invariant and discussed how the discrete symmetries are changed under the noncommutative setup.

The multiplicative renormalization has been applied to the NC scalar QED so that the respective counterterm Lagrangian was obtained. Moreover, the Slavnov-Taylor identities were used in order to show a series of identities relating all the renormalization constants of the theory. These identities are valuable since they allow us to determine the remaining constants from the knowledge of some renormalization constants without further calculation. To conclude the section, we wrote the renormalized group equation for the two-point function for the gauge field in order to introduce the $\beta$ function for the gauge coupling and the $\gamma$ function for the dynamical fields.

 After establishing the renormalization of the model, we proceeded to compute the one-loop-order radiative corrections to the basic functions. We have chosen to compute the simpler radiative contributions that allowed us to determine the full set of renormalization constants:
 gauge, charged scalar and ghost fields self-energies, three- and four-point vertex functions $\left<\phi^\dag \phi A\right>$, $\left<\phi^\dag \phi A A\right>$  and $\left<\phi^\dag \phi\phi^\dag \phi\right>$, respectively. We called special attention to those graphs that were exclusively from noncommutative nature, going to zero at the commutative limit; in particular, we discussed in detail the UV/IR mixing in the respective self-energy functions. Only the counterterm $\delta_1$ related to the vertex $\left<\phi^\dag \phi A\right>$ had a null value (finite contribution only), all the other counterterms absorbed the respective divergent part of the self-energy and vertex functions.
 It should be emphasized that, based on the obtained results, the remaining renormalization constants $Z_{_{3A}}$, $Z_{_{4A}}$, and $\widetilde{Z}_{_{1}}$ can immediately be determined by means of the gauge identities \eqref{STidentity}.

To conclude, we computed the $\beta$ function for the gauge coupling and scalar self-coupling and the anomalous dimensions for the dynamical fields $(A_\mu,\phi,\bar{c},c)$. To highlight the obtained result for the $\beta(e)$, we compared it with the NC-QED and commutative SYM $\beta$ function expressions; as a matter of fact, the gauge sector is strictly the same, and differences are only found in the matter sector contribution.

\subsection*{Acknowledgments}
The authors would like to thank the anonymous referee for his/her comments and suggestions to improve this
paper. We would like to thank M. Khorrami and M. Arjang for useful discussion. R. B. thankfully acknowledges CNPq for partial support, Project No. 304241/2016-4.


\end{document}